\documentstyle[aps,epsfig,amssymb,amsfonts]{revtex}
\tightenlines
\begin{document}
\title{Orbital Currents in Underdoped Cuprates}
\author{Patrick A. Lee}
\address{Center for Materials Science and
Department of Physics, MIT, \\Cambridge, MA 02139}
\date{\today}
\maketitle

\smallskip
\begin{center}
{\it Proceedings of the SNS 2001 Conference}
\end{center}

\begin{abstract}
Orbital currents, either fluctuating or static, have emerged as promising candidates for a description of the
pseudogap state in underdoped cuprates.  I shall review the evolution of these ideas and describe some
experiments which have been proposed to test the existence of orbital currents.
\end{abstract}
\vspace{.40in}

By now there is broad agreement that the physics of high T$_c$ superconductivity is fundamentally the
physics of doping into a Mott insulator.\cite{1}  In this problem the key competition is between the exchange
energy $J$ per spin and the kinetic energy $t$ per hole, i.e., $xt$ per site, where $x$ is the  hole
concentration.  For $xt \gg J$ the kinetic energy wins and the ground state is a Fermi liquid
metal.  For small doping, the N\'{e}el state is destroyed beyond 3\% holes.  The range of doping between
3\% and 15\% is referred to as the underdoped region, where the physical properties are most 
anomalous.  The phenomonology of the pseudogap is well established in this doping range and we would like
to focus our attention on this region.

Recently, Chakravarty {\it et al.}\cite{2} proposed that the pseudogap state is characterized by a hidden order
parameter, which they called the $D$-density wave (DDW).  The physical manifestation of this order parameter is
the appearance of staggered orbital currents.  As Chakravarty {\it et al.} pointed out, this kind of ordered state
is not a new idea and has been proposed by Affleck and Marston\cite{3} and by Schulz\cite{4} in the context of the
$t$-$J$ and the Hubbard model. Indeed, Hsu, Marston and Affleck\cite{5} explicitly proposed this state (called the
staggered flux state) as a description of the underdoped cuprates a decade ago.  In the intervening years, X.-G.
Wen and I and collaborators\cite{6,7} have proposed a fluctuation version where the pseudogap state is understood
not as a genuine phase transition, but as a cross-over to a state with strong fluctuations between the staggered
flux state and the $d$-wave superconductor, and many states smoothly connected to them.  In  this talk I will
briefly review this history and explain why we favored the fluctuation rather than the ordered version of
staggered flux/DDW.  We will end with some proposed experimental tests of our ideas.

We begin with the slave boson formulation of the $t$-$J$ model, where the electron operator
$c_\sigma$ is written as
$c_\sigma = f_\sigma b^\dagger $ with the constraint
$f_\sigma^\dagger f_\sigma + b^\dagger b = 1$ on each site. The exchange interaction can be written in
terms of the fermions as

\begin{eqnarray}
J\left(\stackrel\rightarrow{S}_i \cdot \stackrel\rightarrow{S}_j - \frac{1}{4} n_in_j \right)
	&=& -\frac{J}{2}  \left| f_{i\alpha}^\dagger f_{j\alpha} \right|^2 \\
 &=& -\frac{1}{2} J \left( f_{i\uparrow}^\dagger  f_{j\downarrow} ^\dagger -
f_{i\downarrow}^\dagger f_{j\uparrow}^\dagger 
\right)
\left( f_{j\downarrow}f_{i\uparrow} - f_{j\uparrow}f_{i\downarrow}
\right) 
\end{eqnarray}
These invite the new decoupling schemes\cite{8} (as opposed to the conventional decoupling $\langle S_i
\rangle \cdot S_j$),

\begin{eqnarray}
\chi_{ij} &=& \langle f_{i\alpha}^\dagger f_{j_\alpha} \rangle \\
\mbox{\rm and} \rule{.25in}{0in}
\Delta_{ij} &=& \langle f_{i\uparrow} f_{j\downarrow} - f_{i\downarrow} f_{j\uparrow} \rangle
\end{eqnarray}
It turns out that the best pairing state is a $d$-wave state.  At $x=0$, the best hopping state is one where
the hopping $\chi_{ij}$ is complex, in such a way that the total phase around a plaquette (representing an
effective flux $\phi$) takes the value $\phi=\pi$.\cite{3}  Away from $x=0$, a competitive state is the
staggered flux state, where $\phi$ becomes less than $\pi$ and takes on a staggered pattern.\cite{3}  This
state doubles the unit cell, with the remarkable property that the fermion dispersion is identical to that of
the $d$-wave state, with zero chemical potential and $|\Delta| / |\chi | = \tan (\phi / 4)$ .  This
dispersion is characterized by Dirac fermions centered around nodes at $(\pm\pi/2, \pm\pi/2)$.  The origin
of the degeneracy at $x = 0$ was identified as SU(2) symmetry where the fermions form an SU(2) doublet 
$\Psi_\uparrow = \left(  
\begin{array}{c}
f_\uparrow \\
f_\downarrow ^\dagger
\end{array}
\right)
$.\cite{9}
The Lagrangian was shown to be symmetric under SU(2) rotation at half filling.  Thus the $\pi$ flux state
and the $d$-wave state (with $|\Delta | = |\chi |$) describes the same physical state once the constraint is
imposed.  With doping it was shown that the $d$-wave state is favored in mean field theory over the
staggered flux state.\cite{10,11}  The reason is that in the pairing state, the energy gap structure is pinned to
the Fermi surface, whereas in the staggered flux state, a small pocket of holes appears around the nodal
points at $(\pm \pi /2, \pm \pi /2)$, which costs kinetic energy of order $J x^{3/2}$.  (This is because
with the linear density of states of the Dirac spectrum, the chemical potential $\mu \approx x^{1/2}$.) 
Thus in the mean-field theory (which we shall refer to as the U(1) mean field theory), the pseudogap state
was identified as the fermion pairing state.\cite{10,11,12}

It is interesting to note how much of the pseudogap phenomenology was predicted by the early slave boson mean
field theory.  This includes the appearance of an energy gap in the spin excitations and in the $c$-axis
conductivity, the $d$-wave nature of the photoemssion gap, and the absence of a change in the spectral weight for
the $ab$ plane conductivity.  As far as I know, this is the only theory which anticipated the pseudogap
phenomenology, which itself took many years before it was widely accepted by the community.
 In the midst of many uncertainties surrounding both theory and experiment, in 1991 Hsu, Marston and
Affleck\cite{5} made a serious proposal for the staggered flux state as a candidate for the pseudogap phase in
underdoped cuprates.  They pointed out that with doping, an orbital current (with magnitude proportional to $x$)
will flow around each plaquette in a staggered pattern.  These currents generate a small magnetic field, also
forming  a staggered pattern, and they estimated the field strength just above the plaquette center to be about 10
gauss.  They pointed out that a
$\mu$-SR experiment may be able to detect this magnetic field. They also calculated the neutron scattering
cross-section due to the orbital currents and concluded that they lead to Bragg scattering at $(\pi,\pi)$
with an intensity approximately 1\% of that of the insulator, i.e., with an effective moment of 0.1 Bohr
magneton.  These proposals were largely ignored at the time, partly because the phenomenology of the
pseudogap was not yet fully appreciated, and partly because the staggered flux state was energetically
unfavorable compared with the $d$-wave pairing state.\cite{10,13}

In 1996, Wen and Lee\cite{6} introduced a new formulation of the $t$-$J$ model, where the SU(2)
symmetry is preserved even away from half filling.  The price we pay is that instead of a single slave boson,
a pair of boson fields $(b_1,b_2)$ forming an SU(2) doublet are needed on every site.  The SU(2) singlets
formed out of the fermion and boson singlets and doublets represent the three physical states of singly
occuped and empty states.  While formally the SU(2) and U(1) formulation are equally exact, they lead to
different mean-field theories.  In particular, while the U(1) mean-field theory picks out the $d$-wave pairing
state and completely ignores the staggered flux state, the SU(2) mean-field theory includes fluctuations
between them.  Since their energy difference goes to zero as $x^{3/2}$ for small $x$ as mentioned earlier,
Wen and Lee argued that these fluctuations are important in the underdoped region.  A later reformulation
emphasized these fluctuations by assuming that the fermions follow a locally Bose condensed bosonic
coordinate.\cite{7}  In this $\sigma$-model formulation, the fluctuations between different U(1) mean-field
states, such as the $d$-wave superconductor and the staggered flux state, are made explicit.  The
pseudogap region is described as a state which fluctuates in space and time between $d$-wave
superconductivity and staggered flux, and many states continuously connected in between.  There exists
only short-range order in both order  parameters.  However, since the energy gap at $(0,\pi)$ is common to
both states, the energy gap is a robust feature of this fluctuating state.  These fluctuations give rise to
low-lying electron excitations near the nodal region $(\pi /2,\pi /2)$.  By including a residual interaction
between fermions and bosons, the Fermi surface segments observed by ARPES can be explained.\cite{14}

The above work was based largely on mean-field theory, where the constraint was imposed only on
average. It should be noted that the mean-field theory must be supplemented by gauge fluctuation in order to make
physical sense.  The fermions and bosons are not physical objects, and neither are order parameters such as
$\chi_{ij}$ and $\Delta_{ij}$.  One way of seeing this is the physical charge of the electron can be attached to
the boson or the fermion, or by any arbitrary amount, as long as the net charge adds up to one.  The minimal
theory must include gauge fluctuation at the Gaussian level, and it was shown by Ioffe and Larkin \cite{15}, and
worked out explicitly by Sachdev,\cite{16} that once gauge fluctuations are included, physical quantities such as
conductivity do not depend on the partition of charges between fermion and boson.  Thus the physical consequences
of the mean-field solution such as the appearance of the spin gap are robust, even though the order parameters are
not and the mean-field phase transitions are merely cross-overs.  The only genuine transition is the onset of
superconductivity.  This being said, it is also true that the  coupling to the gauge field is strong.  It can be
controlled formally by a $1/N$ expansion, in which case the coupling is of order $1/N$ after being screened by high
energy fermionic excitations but in the realistic case a coupling of order unity is expected.  Furthermore,
nonperturbative effects such as instantons are known to be important in pure gauge theories and lead to
confinement.  It is possible that instantons are not important in the presence of matter fields (finite doping),
but this cannot be proved or disproved.  Our best hope is to explore predictions of physical quantities and look to
experimentalists to prove or disprove this line of approach.  Already we saw that the theory in its simplest form
had predictive power as far as the pseudogap is concerned.  Recently, we have explored the issue of orbital
current fluctuations which we believe to be an integral part of the gauge theory description of the underdoped
$t$-$J$ model.  We found encouraging support from the study of projected wavefunctions and from exact
diagonalization of small samples.  We report on these below and go on to make predictions for experimental tests.

One way of imposing the constraint exactly is to perform a Gutzwiller projection (i.e., imposing the
no-double-occupancy constraint by hand on a computer) on the mean-field state.  The Gutzwiller projected
$d$-wave BCS superconductor state has been studied for over ten years and shown to be a good trial
wavefunction  for the doped $t$-$J$ model.\cite{17}  Ivanov {\it et al}.\cite{18} took this well known
wavefunction and computed the equal time current-current correlation function $C_J = \langle j_\alpha
j_\beta \rangle$, where $j_\alpha= it \left( c_{i\sigma}^\dagger c_{j\sigma} - c_{j\sigma}^\dagger
c_{i\sigma} \right)$ is the electron orbital current operator on the bond $\alpha$ which connects the
nearest-neighbor sites $i$ and $j$.  Since the wavefunction is translationally invariant,  $\langle j_\alpha 
\rangle = 0$.   Remarkably, the correlator $C_J$ exhibits current loops in  a staggered pattern, which
decays relatively slowly in a power law manner.  Note that these patterns are absent in the BCS state, and
emerge only after Gutzwiller projection.  This is in agreement with the expectation of SU(2) mean-field
theory, in that staggered flux fluctuations are predicted to be  important even in the $d$-wave
superconducting ground state.  Soon after this work, short-range staggered orbital current fluctuations
were discovered in the exact ground state of a small $t$-$J$ system (2 holes in 32 sites.)\cite{19} 
These results are very encouraging for the SU(2) theory,and greatly strengthen the notion that the
pseudogap state is characterized by strong staggered-flux fluctuations.  Of course, the ultimate test  of
these ideas lies in experiments, and in the following we will suggest a number of experimental tests.

It should be pointed out that in our view the orbital current is a ``symptom'' of the state which has the
advantage that it is perhaps more amenable to detection.  The more fundamental driving force lies in the
chirality of the spin arrangement and its effect on the hole motion.  Consider four sites around a plaquette
with  three spins and a hole.  At a given instance the quantization axes of the spins subtend a solid angle
$\Omega$.  When the hole hops around a plaquette, the effect of the changing quantization axes leads to a
Berry's  phase $\Omega /2$ for the hopping amplitude.\cite{20,21}  Thus the hole experiences an
effective gauge flux $\phi = \Omega /2$ and the hole prefers to travel clockwise or counter-clockwise
depending on the sign of the flux $\phi$ (or, equivalently, depending on whether the solid angle is less than
or greater than $2 \pi$).  The appearance of non-coplanar spin arrangments (flux or spin chirality) provides
an improved compromise between the exchange energy of the spin and the kinetic energy of the hole,
compared with the collinear arrangement of the N\'{e}el state.  The staggered arrangment of the flux (or
chirality) is quite natural, given the antiferromagnetic nature of the exchange.  We have tested this point of
view by computing the equal time correlation of the following operator for sites 1 through 4 around a
plaquette
\begin{displaymath}
\chi_h = \left(
\stackrel\rightarrow {S}(1) \cdot \stackrel\rightarrow {S}(2) \times \stackrel\rightarrow {S}(3)
\right)
n_h (4)
\end{displaymath}
where $n_h = 1-c^\dagger c$ is the hole number.\cite{22}   This correlation shows the same power law
decay we found for the current-current correlation.

Recently, the idea that the pseudogap phase may be characterized by staggered orbital currents was revived
by Chakravarty {\it et al}.\cite{2}  The physical nature of the long-ranger order, ie., that the unit cell  is
doubled by the appearance of staggered orbital currents,\cite{23} is identical to that proposed by Hsu {\it
et al}.
\begin{eqnarray}
\langle c_j^\dagger c_{j+\hat{x}} \rangle &=& iJ_0 (-1)^{j_x+ j_y} \\
\langle c_j^\dagger c_{j+\hat{y}} \rangle &=& -i J_0 (-1)^{j_x+j_y} \,\,\, .
\end{eqnarray}
The key point is that this amplitude is imaginary, leading to current modulations without density
modulations.  The Fourier transform of Eqs. (5) and (6) gives

\begin{equation}
\langle c_{\stackrel\rightarrow{k},\sigma}^\dagger c_{\stackrel\rightarrow{k}+
\stackrel\rightarrow{Q} _{0,\sigma}} \rangle = iJ_0 \left( \cos k_x - \cos k_y \right)
\end{equation}
where $Q_0 = (\pi,\pi)$ and

\begin{equation}
y = i \sum_{k,\sigma} \left( \cos k_x - \cos k_y \right)
\langle c_{\stackrel\rightarrow{k}\sigma}^\dagger c_{\stackrel\rightarrow{k}+\stackrel\rightarrow{Q}
_{0,\sigma}}  \rangle
\end{equation}
is non-zero.  Chakravarty {\it et al}. refer to $y$ as the order parameter which they call $D$-density wave
(DDW).  Compared to the earlier work, their approach is more phenomenological in that the existence of the
long-range order is simply assumed and then its consquences are explored using Landau theory.  They
motivate their choice of order by a Hartree-Foch analysis of what looks like a $t$-$J$ model but without any
constraint of no-double-occupancy.  In this sense their motivation is closer in spirit to the work of
Schulz,\cite{4} who analyzed a weak coupling Hubbard model and proposed the same kind or ordering. 
Unless one abandons the basic premise that the high T$_c$ problem is one of a doped Mott insulator, ie.,
strong correlation, it is difficult to see how this weak coupling analysis can motivate the choice of the order
parameter. 
 
Even as pure phenomenology, we find the argument for an ordered DDW state unconvincing.  The existence
of long-range order implies a definite transition temperature, where a specific heat peak should be observed. 
No such peak has been reported.  Chakravarty {\it et al}. argued that disorder is responsible for smearing
this transition.    They argue that the DDW transition is in the same universality class as the
random bond Ising model.  At the same time, they correctly pointed out that weak disorder is irrelevant
and cannot destroy the long-range order of the phase transition.  Indeed, in order to destroy the long-range
order, it must be energetically favorable to introduce domain wall and it is intuitively obvious that some bonds
must change sign before this can happen.  Thus we conclude that in order to suppress the transition, the
disorder must be strong enough to change the sign of the local order.  However, in the present case, potential
fluctuations due to variations in the dopant distribution do not couple directly to the DDW order parameter
because it involves current modulation but no density modulation.  Local modulations in the hole density may
suppress the current locally.  However, large modulations will be required to reverse the direction  of the local
current.  While large density modulations are observed on BSCCO surfaces cleaved at low
temperatures,\cite{24} such variations are not consistent with the micron-scale
quasiparticle mean-free path observed in good quality YBCO samples.\cite{25}  Thus we conclude that
disorder in YBCO samples is not sufficient to destroy the DDW transition if it exists.  Only defects such as
dislocations directly couple to the DDW order parameter, but this kind of disorder is rare in the high quality
samples being studied  today.

A second difficulty of the DDW phenomenology, which is expressed in terms of electron coordinates, is that
the ordered state should be a Fermi liquid, with small pockets of area $x$.  Such pockets have not been seen
by ARPES measurements, and transport properties such as the Hall effect do not fit a Fermi liquid description
of small Fermi pockets.  Finally, a phenomenology based on conventional gapping of a Fermi surface always
leads to a reduction of the spectral weight of the Drude peak in the optical  conductivity.  Chakravarty {\it et
al}. in fact  appeal to such a reduction to explain  the small superfluid density in the underdoped region.  It then
follows that the same picture must predict a restoration of the full spectral weight of $1-x$ electron per site
above the DDW transition.  No such spectral weight reduction has been seen  with the onset of the
pseudogap.\cite{26,27}

Thus we continue to favor our picture of fluctuating orbital currents which was designed to circumvent the
difficulties associated with an ordered state. We further emphasize that spectral weight issues $\left(x
\mbox { \rm vs } (1-x)\right)$ are inherent to the doped Mott insulator problem.  These questions are
conveniently addressed using slave bosons and gauge fields, but not in terms of the original electrons.
 Of course, it is the experimentalists who will have the final
say.  At the moment, two kinds of experiments appear relevant to this issue.  The first is neutron
scattering, which reports the appearance of weak magnetic Bragg scattering at $(\pi,\pi)$ below about
300~K in  two  underdoped samples.  In the experiment of Sidis {\it et al}.,\cite{28} the magnetic order was
identified as due to spin lying in  the plane, just as in the insulator, but with a scattering intensity a factor of
a hundred smaller.  While the intensity is consistent with the estimate of Hsu {\it et al}.,\cite{5} the form
factor and polarization dependence clearly rule out an orbital origin.  At the moment the source of this
scattering is quite a mystery, but small pieces (1\% in  total volume) of some insulating N\'{e}el state cannot
be definitely ruled out.  Sidis {\it et al}. also report additional scattering which onsets below the
superconducting $T_c$.  The linewidth is quite different from the high temperature peak, suggesting that it
could be of different origin.  At present there is no form factor information on  this additional scattering,
so it remains a candidate for orbital contribution.  Mook {\it et  al}.\cite{29} also observed scattering which
onsets at about 300~K, but their intensity is another factor of 10 smaller.  Their form factor analysis
suggests that it is not due to spin lying in the plane, but given the small size of the signal, no definitive
assignment was made.  This is clearly an  important experiment which merits further investigation.\cite{30} 
The second class of experiments is $\mu$-SR, which were done on two samples with higher hole doping
than the neutron samples.\cite{31}  The experimentalists report the onset of some slowly fluctuating 
magnetic fields below
$T_{onset}$.  However, it is impossible to distinguish between spin freezing and the onset of orbital currents
by this experiment.  Thus we consider the experimental evidence for DDW  order to be inconclusive at present.

Fluctuating orbital current loops give rise to dynamical magnetic fields which are, in principle, detectable by
neutron scattering.  In practice the small signal makes this a difficult, though not impossible experiment and we
are motivated to look for situations where the orbital current may become static or quasi-static.  Recently, we
analyzed the structure of the $hc/2e$ vortex in the superconducting state within the SU(2) theory and concluded
that in the vicinity of the vortex core, the orbital current becomes quasi-static, with a time scale determined by
the tunnelling between two degenerate staggered flux states.\cite{32}  It is very likely that this time is long on
the neutron time scale.   Thus we propose that a quasi-static peak at  $(\pi,\pi)$ will appear in neutron
scattering in a magnetic field, with intensity proportional to the number of vortices.  The time scale may
actually be long enough for the small magnetic fields  generated by the orbital currents to be detectable by
$\mu$-SR or Yttrium NMR.  Again, the signal should be proportional to the external fields.  (The NMR experiment
must be carried out in 2--4--7 or 3 layer samples to avoid the cancellation between bi-layers.)  We have also
computed the tunnelling density of states in the vicinity of the vortex core, and predicted a rather specific kind
of period doubling which should be detectable by atomic resolution STM.\cite{33}  

Our picture of the vortex core implies that when $H$ exceeds $H_{c2}$, the vortex core overlaps and the
sample should be described by an ordered staggered flux state.\cite{32}  This is a fermi liquid ground state
with unit cell doubling and small Fermi pockets. Many experimental tools, such as cyclotron resonance,
deHaas-van Alphen, etc., are available to test the unique band structure of the staggered flux state.  Recent
reports of highly underdoped, high quality samples with $T_c$ below 10~K have greatly improved the
prospects for this class of experiments.\cite{34}

Finally, we address the question of why it is so important to have a state other than superconductivity to
describe the pseudogap region.  One view of the pseudogap phase is that it is a local superconductor with robust
amplitude but strong phase fluctuations.  That this view is incomplete can be seen from the following argument.  In
two dimensions the destruction of superconducting order is via the Berezinskii-Kosterlitz-Thouless (BKT) theory of
vortex unbinding.  Above T$_c$ the number of vortices proliferate and the normal metallic state is reached only
when the vortex density is so high that the cores overlap.  (There is considerable lattitude in specifying the core
radius, but this does not affect the conclusion.)  At lower vortex density, transport properties will resemble a
superconductor in the flux flow regime.  In ordinary superconductors, the BKT temperature is close to the mean
field temperature, and the core energy rapidly becomes small.  However, in the present case, it is postulated that
the mean field temperature is high, so that a large core energy is expected.  Indeed, in a conventional core, the
order parameter and energy gap vanish with an energy cost of $\Delta_0^2/E_F$ per unit area.  Using a core radius
of $\xi = V_F/\Delta_0$, the core energy of a conventional superconductor is $E_F$.  In our case, we may replace
$E_F$ by $J$.  If this were the case, the proliferation of vortices will not happen until a high temperature
$\sim J$ independent of $x$ is reached.  Thus for the phase fluctuation scenario to work, it is essential to
have ``cheap'' vortices, with energy cost of order  T$_c$.  Then the essential problem is to understand what the
vortex core is made of.  In the SU(2) theory we have a natural candidate, the staggered flux phase, and we have
successfully constructed a ``cheap'' vortex state.\cite{32}  Other possibilities, such as a  N\'{e}el ordered
state or spin density waves,\cite{35,36,37} have been proposed.  In our view, the staggered flux phase has an
advantage over other possibilities in that its excitation spectrum is similar to the $d$-wave superconductor.  In
any event the theory is not complete until the nature of the alternative state which constitutes the vortex core is
understood.  Then the pseudogap phase can be understood equally well as fluctuating superconductors with
regions of the alternative state or as a fluctuating alternative state with regions of superconductivity.

\acknowledgements{I acknowledge close collaboration  with X.-G. Wen on the development of the SU(2)
theory, and I  thank D. Ivanov, J. Kishine,  N. Nagaosa, and  T.K.  Ng  for their collaboration and discussions. 
I also thank T. Senthil for helpful comments.  This work was supported by the NSF through the MRSEC
Program DMR 98--08941.}


\begin{thebibliography}{99}
\bibitem{23}
Note that this is fundamentally different from the orbital current proposed by C.M. Varma, Phys. Rev. Lett. 
{\bf 83}, 3538 (1999) which is within the unit cell and does not lead to unit cell doubling.
\end{thebibliography}
\end{document}